\documentclass[twocolumn,longbibliography,aps,prl,superscriptaddress,showpacs,floatfix]{revtex4-1}
\usepackage{graphics,epsfig,graphicx}
\usepackage{amsbsy,gensymb}
\usepackage{amsmath}
\usepackage{bm}
\usepackage{amsfonts}
\usepackage{cancel}
\usepackage{multirow}
\usepackage{xcolor}
\begin{document}

\title{From few to many bosons inside the unitary window: a transition between
universal to non-universal behavior}

\author{A. Kievsky} 
\affiliation{Istituto Nazionale di Fisica Nucleare, Largo Pontecorvo 3, 56100 Pisa, Italy}
\author{A. Polls} 
\author{B. Juli\'a-D\'\i az}
\affiliation{Departament de F\'\i sica Qu\`antica i Astrof\'\i sica, Facultat de F\'\i sica,
Universitat de Barcelona, E-08028 Barcelona, Spain}
\author{N. K. Timofeyuk}
\affiliation{Department of Physics, University of Surrey, Guildford, Surrey GU2 7XH, United Kingdom}
\author{M. Gattobigio}
\affiliation{
 Universit\'e C\^ote d'Azur, CNRS, Institut  de  Physique  de  Nice,
1361 route des Lucioles, 06560 Valbonne, France }

\begin{abstract}
Universal behaviour in few-bosons systems close to the unitary limit, where two bosons become 
unbound, has been intensively investigated in recent years both experimentally and theoretically. 
In this particular region, called the unitary window, details of the inter-particle interactions are
not important and observables, such as binding energies, can be characterized by a few parameters. 
With an increasing number of particles the short-range repulsion, present in all atomic, molecular 
or nuclear interactions, gradually induces deviations from the universal behaviour. In the present 
letter we discuss for the first time a simple way of incorporating non-universal behaviour through 
one specific parameter which controls the smooth transition of the system from universal to non-universal
regime. Using a system of $N$ helium atoms as an example we calculate their ground state energies as 
trajectories within the unitary window and also show that the control parameters can be used to 
determine the energy per particle in homogeneous systems when $N \rightarrow \infty$.
\end{abstract}

\maketitle

{\bf Introduction.}
Close to the unitary limit, the physical behaviour and properties of few-body systems are driven and shaped 
by universality and this has far-reaching consequences for $N$-particle systems.
At this limit a two-body system has a bound state at its decay
threshold, with the two particles staying mostly outside the region of their interaction.
The properties of this system are determined by one parameter, the two-body energy length $a_B$, 
defined from the two-body binding ($a_B>0$) or virtual ($a_B<0$) energy 
$E_2=\hbar^2/m a_B^2$ ($m$ is the particle mass). In the limit of a zero-range 
interaction, the two-body scattering length $a$ and the energy length are equal, 
$a=a_B$, and the two-body system shows a continuous scale invariance. For finite 
range interactions $a \neq a_B$ and the difference  $r_B=a-a_B$, called the finite-range parameter,
defines the  unitary window if the condition $r_B/a_B\approx r_B/a \ll 1$ is satisfied.

The special  nature 
of the unitary window shows up
in a dramatic way in the energy spectrum of three-body systems, as shown by V. Efimov
for the case of a zero-range attractive interaction~\cite{efimov1,efimov2}.
The system has a discrete scale invariance which is manifest 
at unitarity by the Efimov effect: an infinite tower of geometrically-distributed energy states 
with the neighbouring energies ratios of $\approx 515$. 
Intense experimental efforts, notably in the field of ultracold 
quantum gases~\cite{kraemer2006,zaccanti2009,ferlaino2011,matchey2012,roy2013,cornell2017},
as well as theoretical studies~\cite{report,naidon} have been dedicated to this subject,
including larger systems~\cite{platter,deltuva,greene,kievsky2014,bazak} or those with different
symmetries~\cite{kievskyfbs,koenig,gatto2019b}. 

A large class of systems inside the 
unitary window is well described using a simple gaussian interaction, 
\begin{equation}
 V(r_{ij})= V_0 \,e^{-r_{ij}^2/r_0^2}
\label{eq:tbgaus}
\end{equation}
with a variable strength $V_0$ ($r_{ij}$ is the interparticle distance)~\cite{kievsky2014,raquel}. 
In this way the universal behavior, exactly verified in the case of zero-range interactions,
is extended to include finite-range corrections~\cite{gatto2019}.
Finite range effects become more important when the interparticle distance inside the $N$-boson clusters
gets sufficiently small so that the short-range physics 
starts to manifest explicitly in a non-universal way because of a different repulsive core in each 
particular system.
This effect shows up smoothly with an increasing number of particles  driving the system
from a universal regime to a non-universal one.

In this letter we study the transition to non-universality in a two step analysis. First of all,
we perform a gaussian characterization of the unitary window of the $N$-boson system 
by constructing trajectories in the energy plane $(E_2,E_N)$ using the interaction potential  
of Eq.~(\ref{eq:tbgaus}).
In this plane, any system can be represented by a point, called the physical point,
determined when the $E_2$ and $E_N$ ground state energies are simultaneously reproduced
by the gaussian parameters. 
In the case of ultracold atomic gases, tunable Feshbach resonances
can be used to explore experimentally the unitary window~\cite{julienne}. It can
be also explored theoretically by varying the inter-particle potential. The movement of 
the system along the path determined by the gaussian form reveals its universal character. 

The second step of our analysis uses the effective field theory (EFT) framework 
introduced to describe boson systems with large two-body scattering lengths~\cite{vankolck1,vankolck2}.
In this formalism the potential in Eq.~(\ref{eq:tbgaus}) 
enters at leading order (LO); at the same order a three-body force is needed to counterbalance
the dependence introduced by the gaussian range $r_0$. 
The strengths of the two- and three-body LO terms 
are determined by two control parameters, $E_2$ and $E_3$. 
In a universal regime the energies $E_N$, $N>3$, are completely determined
by the two control parameters, except for a residual range dependence~\cite{carlson}. 
We explore this dependence and show that the range of the three-body force, that 
could differ from the two-body range $r_0$,  emerges as a non-universal
scale parameter useful to describe the $N$-boson systems 
inside the window. Using helium systems as an example, by setting this parameter to describe 
$E_4$ together with $E_2$ and $E_3$, we show that energies per particle, $E_N/N$, as 
$N\rightarrow\infty$ can be well reproduced.

\begin{figure}[h]
\vspace{-25pt}
\includegraphics[scale=0.34]{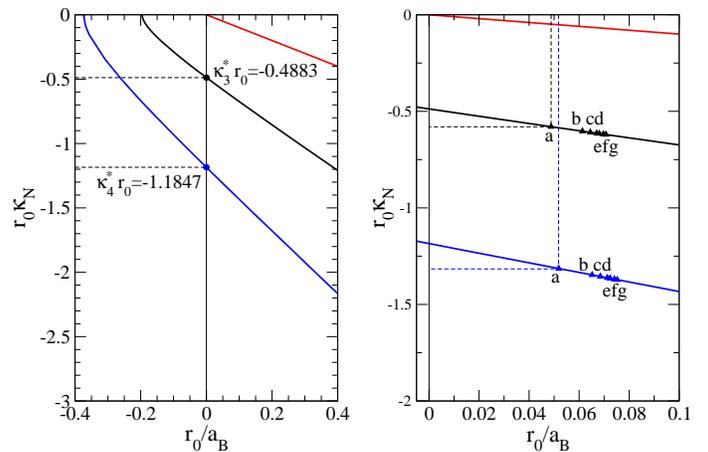}
\caption{Binding momentum in terms of the inverse of $a_B$ for a gaussian
potential, in units of the gaussian range $r_0$. The cases $N=2,3,4$ are shown by
 the red, black, and blue solid lines, respectively (left panel). Specific helium trimers
and tetramers are located on the plot, see text for details. As an 
example, the dashed lines mark the location of the ``a'' point.}
\label{fig:fig1}
\end{figure}

{\bf Gaussian characterization for $N$ bosons.}
The ground state energies of $N=2,3,4$ bosons along the unitary window, obtained using the
gaussian interaction of Eq.~(\ref{eq:tbgaus}), are represented
in Fig.~\ref{fig:fig1} through their binding momenta $\kappa_N$, defined from $E_N=\hbar^2\kappa^2_N/m$.
They are plotted as functions of the inverse of $a_B$ and all
quantities are made dimensionless by being scaled by  the gaussian range $r_0$. 
The figure relates few- and two-body energies in a unique way: 
gaussians with different ranges and strengths give results that always 
lie on the same curves. At unitarity the
quantities $\kappa^*_N r_0=0.4883$ and $1.1847$, for $N=3,4$, respectively, are the 
same for all gaussian interactions. These points are highlighted in the left panel of 
Fig.~\ref{fig:fig1}.  

Real systems are located on the gaussian plot of Fig.~\ref{fig:fig1}
through the energy ratio $E_N/E_2$. As an example we discuss clusters of He atoms 
which are among a few physical systems naturally existing 
inside the unitary window. Early estimates of the two-body 
scattering length $a\approx 180\, a_0$ and the dimer energy  
$E_2\approx 1\,$mK were given in~\cite{grisenti}. Recently $E_2=1.70 \pm 0.15\,$mK
was measured by Coulomb explosion~\cite{kunitski}.
Due to the relatively large experimental uncertainty in these values we plot results of 
theoretical calculations
for these systems noticing  that a few of them agree with the measured values. 
We consider two-, three- and four-body energies calculated in Ref.~\cite{hiyama2012} 
for a variety of realistic He-He  interactions shown in Table~\ref{tab:xaxis}. 
Using these results we calculate the ratios $E_3/E_2$ and $E_4/E_2$ and display the physical points, 
corresponding to the interactions listed in Table~\ref{tab:xaxis}, in the 
right panel of Fig.~\ref{fig:fig1}. 
A particular gaussian range $r_0$ can be determined
for each He-He potential from the corresponding axis values, $r_0/a_B$ or $r_0\kappa_N$.
Interestingly, the different $r_0/a_B$ (or $r_0\kappa_N$) axis values, associated with
different He-He potentials, correspond to an almost unique value of $r_0$ in each case of $N$: 
$r_0^{(3)}$ for $N=3$ and  $r_0^{(4)}$ for $N=4$, both shown  in Table~\ref{tab:xaxis}.

\begin{table}[ht]
 \begin{tabular} {@{}l c c c c c @{}}
\hline\hline
Potential  & $E_2$  & $E_3$ & $E_4$ & $r_0^{(3)}$($a_0$) & $r_0^{(4)}$($a_0$) \\
\hline
a: HFD-HE2\cite{aziz1}     & $0.8301$  & $117.2$   & $535.6$   & $11.146$ & $11.840$  \\
b: LM2M2\cite{lm2m2}       & $1.3094$  & $126.5$   & $559.2$   & $11.150$ & $11.853$  \\
c: HFD-B3-FCH\cite{hfdb3fch}& $1.4475$  & $129.0$   & $566.1$   & $11.148$ & $11.853$  \\
d: CCSAPT\cite{ccsapt07}    & $1.5643$  & $131.0$   & $571.7$   & $11.149$ & $11.851$  \\
e: PCKLJS\cite{pckljs}      & $1.6154$  & $131.8$   & $573.9$   & $11.148$ & $11.852$  \\
f: HFD-B\cite{hfdb}         & $1.6921$  & $133.1$   & $577.3$   & $11.149$ & $11.854$  \\
g: SAPT96\cite{sapt96}      & $1.7443$  & $134.0$   & $580.0$   & $11.147$ & $11.850$  \\
\hline
\end{tabular}
 \caption{Dimer, trimer and tetramer energies (in mK) for the indicated potential
(the labels indicate the points on Fig.~1).
Values (except for the HFD-HE2 potential) are 
from Ref~.\cite{hiyama2012}. The last two columns
show the $N=3,4$ characteristic gaussian ranges.}
  \label{tab:xaxis}
\end{table}

The fact that all $r_0$  values, determined by different realistic He-He potentials, 
are practically the same for a given $N$ 
allow us to construct the following gaussian potentials
\begin{equation}
 V^{(N)}(r_{ij})= V^{(N)}_0 e^{-r_{ij}^2/(r^{(N)}_0)^2}, \quad {\rm with}\;N=3,4\,.
\label{eq:tbgaus3}
\end{equation}
We will call $r_0^{(N)}$ the characteristic range.
Choosing specific  $V^{(3)}_0$ values, with the average range $r_0^{(3)}=11.147\,a_0$, 
the above potential can reproduce simultaneously the dimer and trimer energies of 
different realistic He-He potentials. Similarly, specific $V^{(4)}_0$ choices can reproduce the 
dimer and tetramer energies. 
The potentials $V^{(3)}$ and $V^{(4)}$ can be
thought of as low energy representations of the realistic interactions.
We will call them characteristic gaussian potentials. 
Decreasing the gaussian strengths 
allows the unitary limit to be reached where the relations $\kappa^*_3r^{(3)}_0=0.4883$ and 
$\kappa^*_4r^{(4)}_0=1.1847$ can be used to calculate the values
%
\begin{eqnarray}
 E^*_3&=83.05 \pm   0.05 \,{\rm mK} \\
 E^*_4&= 433.0 \pm 0.5 \,  {\rm mK} \, .
\end{eqnarray}
%
They should be compared to the values $E^*_3\approx 84.0\,$mK and $E^*_4\approx 439.0\,$mK
obtained for the realistic potentials once their strength is varied to locate 
three- and four-body systems at the unitary limit~\cite{hiyama2014}. 
The quality of the description is around $1\%$ which is a remarkable result. 
The gaussian energy curves coincide with
  those obtained using reduced-depth realistic helium potentials. In
  other words, the characteristic gaussian potentials
  determine a path followed by the realistic systems all the way towards the
  unitarity where the $r_0^{(N)}k_N$ values do not depend on the choice of
  one specific He-He potential. This can be seen as an evidence for universal behaviour.

Next we use the gaussian potential of Eq.~(\ref{eq:tbgaus}) 
to characterise the unitary window for larger number of particles $N$.
Using the hyperspherical harmonic method~\cite{timofeyuk2008,timofeyuk2012},
we calculate the ground state energies for a selected range of $N$ 
and depict the results in Fig.~2.
The energies of the boson systems interacting with the realistic HFD-HE2 potential
are shown in the same figure (solid squares). When this potential
is multiplied by a factor $\lambda$ to reach the unitary limit it gives the results
indicated by the solid circles. We can observe that
at unitarity the energies $E_N$ are on top of the 
gaussian trajectories until $N=10$, suggesting strongly an independence of the interaction details,
and, with small deviations, for $10<N\le 20$. 
Above $N=20$ noticeable differences 
are observed for $N=40$ and $70$ as the short-range physics starts to play a role, resulting in a smooth
transition from a universal to a non-universal regime.

\begin{figure}[t]
\vspace{-25pt}
\includegraphics[scale=0.35]{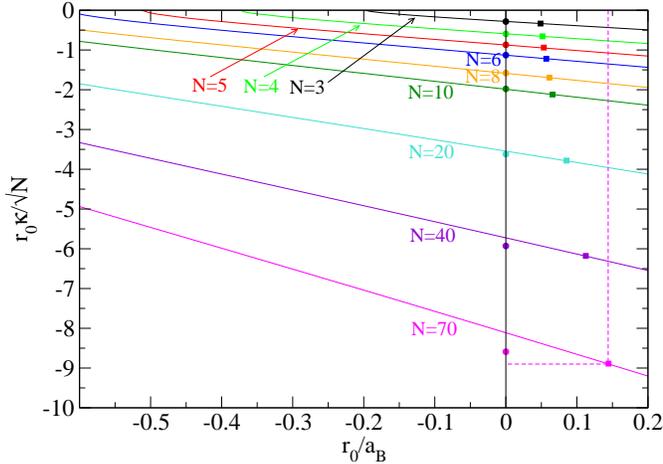}
\caption{The binding momentum per particle $\kappa_N/\sqrt{N}$ in terms of the inverse
of $a_B$ for a gaussian interaction (in units of the gaussian range $r_0$)
and for selected number of particles. For each $N$ value, the lines collect the results of every gaussian potential. The solid squares are the values of the HFD-HE2
potential, the position of each point determines the axis values
$r_0/a_B$ and $r_0 \kappa_N / \sqrt{N}$, as shown by the dashed (magenta) line 
in the $N=70$ case, from which the characteristics range $r^{(N)}_0$ can be determined.
The solid circles at the unitary limit (vertical axis) show the values 
$r^{(N)}_0 \kappa^*_N / \sqrt{N}$ calculated using the HFD-HE2 potential at 
unitarity, i.e., when the potential is multiplied 
by the factor $\lambda=0.9792445$. }
\label{fig:fig2}
\end{figure}

{\bf Soft gaussian potential.}
We have shown above that systems with low values of $N$ display universal behaviour in the unitary window. 
However, the description in terms of the characteristic range, $r^{(N)}_0$,
deteriorates as $N$ increases. To deeper analyse this 
transition we make use of the EFT framework for systems having a large value of the two-body scattering 
length. At LO of this theory~\cite{report,vankolck1,vankolck2} the 
potential consists of a two- plus a three-body term determined to reproduce 
the dimer and trimer energies. We use the following soft gaussian potential (SGP)
\begin{equation}
 V= V_0 \sum_{i<j} e^{-r_{ij}^2/r_0^2} + W_0 \sum_{i<j<k} e^{-2\rho_{ijk}^2/\rho_0^2}
\label{eq:twop3}
\end{equation}
with $\rho^2_{ijk}=(2/3)(r^2_{ij}+r^2_{jk}+r^2_{ki})$.
In the following we use the He-He potential 
HFD-HE2 as a reference potential  
to make a contact with a previous work~\cite{artur}, where saturation properties of 
helium drops were studied from a leading order description.
For $N=2$ (with $\hbar^2/m=43.281307\, {\rm K}a_0^2$), 
this potential gives 
a single bound state, $E_2=0.83012\,$mK, a scattering 
length $a=235.547\,a_0$ and the finite-range parameter $r_B=7.208\,a_0$. For 
$N=3$ and $N=4$ the HFD-HE2 ground state 
energies are given in Table I, obtained using the correlated hyperspherical 
harmonic basis~\cite{barletta} and diffusion Monte Carlo method, 
respectively. They are in good agreement with the Green Function Monte Carlo results of Ref.~\cite{pandha1}.
Reducing the strength of the HFD-HE2 by the factor $\lambda$, 
we decrease the He-He energy down to zero. 
Then we obtain $E_3=83.80\,$mK  and $E_4=439.6\,$mK in close agreement with the results 
of the other realistic potentials.

For a chosen pair of the two- and three-body gaussian ranges $r_0$ and $\rho_0$ we fix the 
SGP strengths $V_0$ and $W_0$ to reproduce the HFD-HE2 energies $E_2$ and $E_3$. Then we use this SGP to 
calculate the tetramer energy  $E_4$.
In Fig.~3 the narrow (green) band shows $E_4$ as a function of $r_0$. The band 
collects the results for different values of $\rho_0$, its lowest part 
is given by the lowest value of $\rho_0$ considered, $\rho_0=3\,a_0$, reducing 
further this value no increase of $E_4$ is obtained. This means that for 
given values of $r_0$ the possible values of $E_4$ are limited and, more 
importantly, only a restricted range  of $r_0$ values is compatible with the energy value 
given by the reference realistic potential and indicated in Fig.~3 by the green horizontal 
line. In the figure the vertical line indicates the $r_0$ value at which both, $E_2$ and
the two-body scattering length $a$, coincide with those of the reference potential HDF-HE2. 
At this particular value, $a-a_B=r_B\approx 7.2\,a_0$, the best description of $E_4$ is obtained.
In Fig.~3 the 
largest value of $r_0$ considered is equal to the characteristic 
range $r^{(3)}_0$, the one that describes the trimer energy 
in the simple two-body gaussian model of Eq.(2).
At this value the three-body force is zero and higher $r_0$ values lead to 
an attractive three-body force not considered in the present analysis.
It should be noticed that in the region limited by $r^{(3)}_0$ and the vertical line
the $E_4$ band is very narrow indicating a low dependence on the three-body range.


\begin{figure}[t]
\vspace{-25pt}
\includegraphics[scale=0.55]{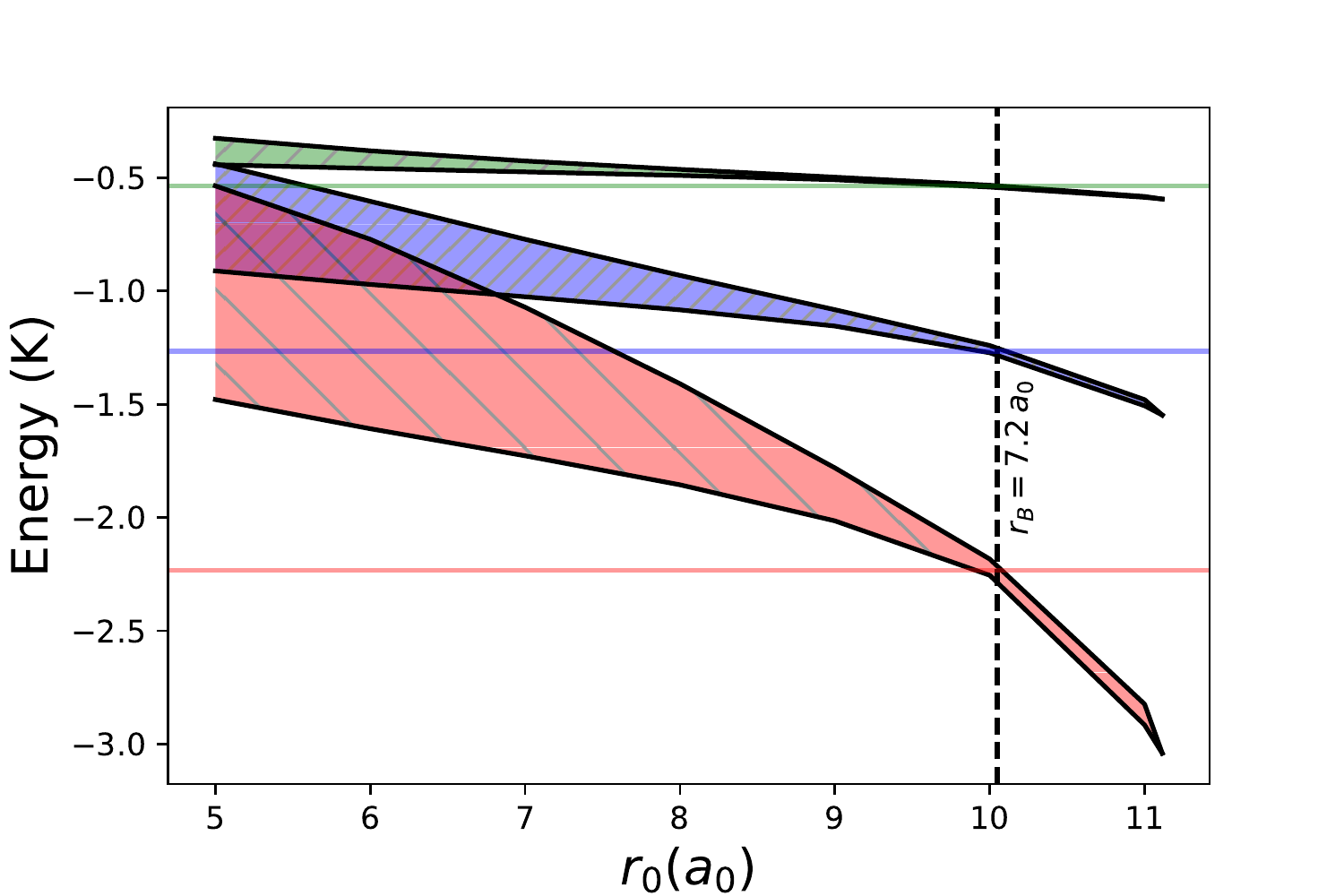}
\caption{$E_4$ (green band), $E_5$ (blue band) and $E_6$ (orange band), as 
functions of the two-body range $r_0$, 
obtained with the three-body 
range, $3\, a_0 \le \rho_0 \le 11\, a_0$. 
The reference 
energies
of the HFD-HE2 potential are given as horizontal lines. The vertical line 
indicates the reference $r_B$ value. Notice 
that the dimer and trimer energies 
 are always reproduced for 
all the SGP interactions considered.
}
\label{fig:fig3}
\end{figure}

Fig.~3 also shows the energy bands obtained for $N=5,6$ systems. In general, they are broader than the one 
corresponding to the $N=4$ case. However, with the SGP parameters reproducing $r_B$ at physical point the 
$N=5,6$  bands become narrow and, more importantly, pass through the  reference HFD-HE2 energies.
A detailed analysis of the results indicates that the best, 
simultaneous, description of $E_5$ and $E_6$ is obtained when the two-body term 
of the SGP potential reproduces the finite range parameter $r_B$ and when the 
three-body range $\rho_0$ is fixed to optimize the description of the tetramer 
energy. The optimum set of these values 
is given in Table~II with the corresponding values of $E_4$-$E_6$
and the HFD-HE2 reference energies, marked as ``physical point''.
In this point the SGP parameters coincide with those of Ref.~\cite{artur}. 
A similar analysis at the unitary point produces the SGP parameters and results given
in the right part of Table~II.

Now we extend our analysis to heavier systems following
a different strategy to the one that has already been used to study few-body systems 
close to the unitary limit at leading order of the EFT ~\cite{bazak}.
There, in order to reduce the residual range dependence of the observables,
the $N\le 6$ binding energies have been studied as $r_0 \rightarrow 0$ and
extrapolated to the zero-range limit $r_0=0$.  Instead, we 
optimize the ranges of the SGP. The two-body range $r_0$ has been
fixed to reproduce two data, $E_2$ and $r_B$ (or equivalently the effective range), 
in order to include finite-range corrections. The resulting two-body potential  
is of the same, next-to-leading, order that potentials 
with two derivatives~\cite{bira}. Furthermore, fixing the three-body range $\rho_0$ 
to optimize $E_4$ we eventually reduce the residual effects of higher order forces. 
The final result is that these four observables, $E_2$, $r_B$, $E_3$ and $E_4$, 
completely determine the SGP.

\begin{table}[t]
 
\begin{tabular} {@{}l  c c | c c  @{}}
\hline\hline
 &  \multicolumn{2}{c}{physical point} &  \multicolumn{2}{c}{unitary point}  \\
\hline
                & SGP      & HFD-HE2 & SGP      & HFD-HE2 \\
$r_0$[$a_0$]    & 10.0485  &         & 10.0485  &  \\
$V_0$[K]        & 1.208018 &         & 1.150485 &  \\
$\rho_0$[$a_0$] & 8.4853   &         & 8.4853   &  \\
$W_0$[K]        & 3.011702 &         & 3.014051 &  \\
\hline
$E_4$[K]        & 0.536    & 0.536   & 0.440    &  0.440  \\
$E_5$[K]        & 1.251    & 1.266   & 1.076    &  1.076  \\
$E_6$[K]        & 2.216    & 2.232   & 1.946    &  1.963  \\
\hline
$E_{10}/10$[K]  & 0.792(2)  & 0.831(2) & 0.714(2)  &  0.746(2)  \\
$E_{20}/20$[K]  & 1.525(2)  & 1.627(2) & 1.389(2)  &  1.491(2)  \\
$E_{40}/40$[K]  & 2.374(2)  & 2.482(2) & 2.170(2)  &  2.308(2) \\
$E_{70}/70$[K]  & 3.07(1)   & 3.14(1)  & 2.80(1)   &  2.92(1)  \\
$E_{112}/112$[K]& 3.58(2)   & 3.63(2)  & 3.30(2)   &  3.40(2)  \\
$E_N/N(\infty)$[K] & 7.2(3)$^*$ & 7.14(2)  & 6.8(3)$^*$ &  6.72(2)  \\
\hline
HFD-B  [K]         &            & 7.33(2)  &            &  6.73(2)  \\
\end{tabular}
 \caption{SGP parameters and the corresponding energies $E_N$ 
 or energies per particle $E_N/N$ at the physical and 
 unitary points. 
 The values indicated with an asterisk ($^*$) are 
 extrapolated results. 
 The energies corresponding to
 the HFD-HE2 potential (in the last row the HFD-B potential) are given too.}
  \label{tab:energy}
\end{table}

In the lower part of Table~II the energy per particle is reported up to $N=112$. 
The results for the infinite system are given in the last two rows where, 
the last one, includes the HFD-B saturation energy. 
We observe that the SGP energies follow  the trend of those obtained with the realistic 
HFD-HE2 interaction which has a strong repulsive core. The weak repulsion in SGP
introduced to describe correctly the trimer is sufficient to guarantee saturation 
of the system. With the selected value of $\rho_0$, the HFD-HE2 energies are
reproduced for all $N$ values within a 5\% accuracy. An extrapolation to the infinite system, 
using a liquid drop formula, maintains the result within this limit (marked with 
an asterisk in the table). This is a remarkable result considering the 
minimal information included in the SGP.

We have discussed the property that different realistic He-He potentials give the same value of $E_3$
and the same value of $E_4$ when their strengths are reduced to locate them at unitarity. So, 
the four observables determining the SGP are independent of the potential used
for its construction and, therefore, the saturation energy predicted by the SGP 
will be the same for all the He-He potentials. This suggests
that all realistic He-He potentials should predict the same saturation 
energy at the unitary limit. To verify this prediction, we have calculated the saturation energy 
for the HFD-B model at the physical and unitary points reported in the last row of Table~II.
Although a difference is observed at the physical 
point, the latter is extremely close to the result of the HFD-HE2 potential 
confirming the collapse to a single value of the saturation energy of the different
He-He interactions at unitarity.

{\bf Conclusions.}
We have shown that the
 universal behavior observed in few-boson systems inside the unitary window
can be characterized by 
paths constructed using gaussian potentials. 
For bosonic helium clusters this behaviour
is well established up to $N\approx 20$ and 
then
smoothly deteriorates for larger $N$ when short-range physics starts to play an explicit role introducing
a non-universal behavior that  competes with the universal characterization of the unitary window. 
Inside the universal regime the gaussian representation explains 
why,  at unitarity, different He-He interactions 
give the same few-body binding energies.

 To map the transition from universal to non-universal regime,
we used the EFT framework, 
introducing a soft gaussian potential having a two-body plus a three-body term.  
Its parametrization, 
constrained from four data
points, 
i.e. the scattering 
length, and the dimer, trimer and tetramer binding energies, 
resulted in a potential that 
predicted reasonably well the $E_N/N$ ratio 
for all $N$, including 
the $N\to\infty$ limit. 
To 
achieve
this unexpected result we performed an optimization of the gaussian ranges, $r_0$ and
$\rho_0$, 
in order
 to reduce effects from higher order terms of the effective expansion that could 
appear in the description of more bound systems. In particular, the
non-universal behavior introduced by the intrinsic 
repulsive
short-range scale was mimicked by 
the properly chosen 
value of $\rho_0$.
Importantly, our characterization can be readily explored in state-of-the-art experiments in 
ultracold quantum gases, where a fine control of the interaction strength is 
achieved allowing a detailed exploration of the unitary window. Finally, let us 
emphasize that our results should be independent of the gaussian form, other representations 
of the zero-range interaction can be used as well with the 
same conclusions~\cite{gatto2014}.

{\bf Acknowledgements.} 
B.J-D and A. P. acknowledge fruitful discussions with A. Sarsa. This work has been partially supported by  MINECO (Spain) Grant No. FIS2017- 87534-P 
and from the European Union Regional Development Fund within the ERDF Operational 
Program of Catalunya (project QUASICAT/QuantumCat). N.K.T. acknowledges support from the United Kingdom Science and Technology Facilities Council (STFC) under Grant No. ST/L005743/1.

\end{document}